\documentclass[pra,twocolumn,showpacs,superscriptaddress]{revtex4-1}
\usepackage[english]{babel}
\usepackage[dvips]{graphicx}
\usepackage{color}
\usepackage{latexsym}
\usepackage{amsmath}
\usepackage{amsthm}
\usepackage{amsfonts}
\usepackage{amssymb}
\usepackage{mathtools}

\newcommand{\bra}[1]{\left\langle #1 \right\rvert}
\newcommand{\ket}[1]{\left\lvert #1 \right\rangle}

\begin{document}

\pagenumbering{arabic}

\title{Decoherence in the quantum Ising model with transverse dissipative interaction in \\ 
the strong coupling regime}

\author{H. Weisbrich}
\affiliation{Fachbereich Physik, Universit{\"a}t Konstanz, D-78457 Konstanz, Germany}
\author{C. Saussol}
\affiliation{Fachbereich Physik, Universit{\"a}t Konstanz, D-78457 Konstanz, Germany}
\author{W. Belzig}
\affiliation{Fachbereich Physik, Universit{\"a}t Konstanz, D-78457 Konstanz, Germany}
\author{G. Rastelli}
\affiliation{Fachbereich Physik, Universit{\"a}t Konstanz, D-78457 Konstanz, Germany}
\affiliation{Zukunftskolleg, Universit{\"a}t Konstanz, D-78457, Konstanz, Germany}

\begin{abstract}
We study the decoherence dynamics of a quantum Ising lattice of finite size with 
a transverse dissipative interaction, namely 
the coupling with the bath is assumed perpendicular
to the direction of the spins interaction and parallel to the external transverse magnetic field.
In the limit of small transverse field, the eigenstates and spectrum are obtained  
by a strong coupling expansion, from which we derive the Lindblad equation 
in the Markovian limit.
At temperature lower than the energy gap and for weak dissipation, the   
decoherence dynamics can be restricted to take only the two degenerate ground states
and the first excited subspace into account.
The latter is formed by pairs of topological excitations (domain walls or kinks), 
which are quantum delocalized along the chain due to the small magnetic field.
We find that some of these excited states form a relaxation-free subspace, 
i.e. they do not decay to the ground states. 
We also discuss the decoherence dynamics for an initial 
state formed by a quantum superposition of the two degenerate ground states 
corresponding to the orthogonal classical, ferromagnetic states.
\end{abstract}
\date{\today}
\maketitle

%
%
%
\section{Introduction}

The quantum (or transverse) Ising model is a unique paradigm  for quantum magnetism and many-body systems
\cite{Lieb:1961,Katsura:1962,McCoy:1968gu,Pfeuty:1970,Barouch:1971gz,Pfeuty:1971ie}. 
It  illustrates the quantum criticality in the one-dimensional quantum phase transition
at  equilibrium \cite{Sachdev:1996eq,Sachdev:1997ht,Suzuki-book:2013,Parkinson-book:2010}.
Several theoretical studies  analyzed  the dynamical aspects of the quantum phase transition \cite{Calabrese:2011eh,Caux:2013gl},  
as the Kibble-Zurek mechanism \cite{Zurek:2005cu,Dziarmaga:2005jq}, 
or  the quantum superposition of topological defects, i.e. domain walls or kinks \cite{Dziarmaga:2011iw}.

Quantum Ising model has been experimentally implemented in artificial quantum many body
systems, as in neutral atoms in optical lattices \cite{Simon:2011ep},
trapped ions \cite{Friedenauer:2008ii,Islam:2011ct,Jurcevic:2017be}
and with arrays  of Rydberg atoms \cite{Labuhn:2016dp}.
It has also been realized more recently in superconducting qubits \cite{Salathe:2015bx,Barends:2016ju,Li:2014}.
In these realizations, the interaction of the system with the environment cannot be disregarded 
since the real chain of effective spins always is affected by a certain amount of dissipation such that 
they have to be considered effectively as open quantum systems \cite{Weiss:2012,Breuer:2012}.

Numerical quantum Montecarlo simulations 
were employed to unveil the phase diagram of a dissipative quantum Ising 
lattice with Ohmic dissipation \cite{Werner:2005bf,Werner:2005,Cugliandolo:2005,Sperstad:2010}.
Effects of the disorder  were  analyzed using a renormalization group approach \cite{Hoyos:2012fh}.    
The phase diagram of a dissipative  Ising model was also recently investigated in the framework of the Lindblad equation 
using a variational approach \cite{Overbeck:2017cw}.

%
%
%
%
%
\begin{figure}[t]
\includegraphics[width=0.9\linewidth,angle=0.]{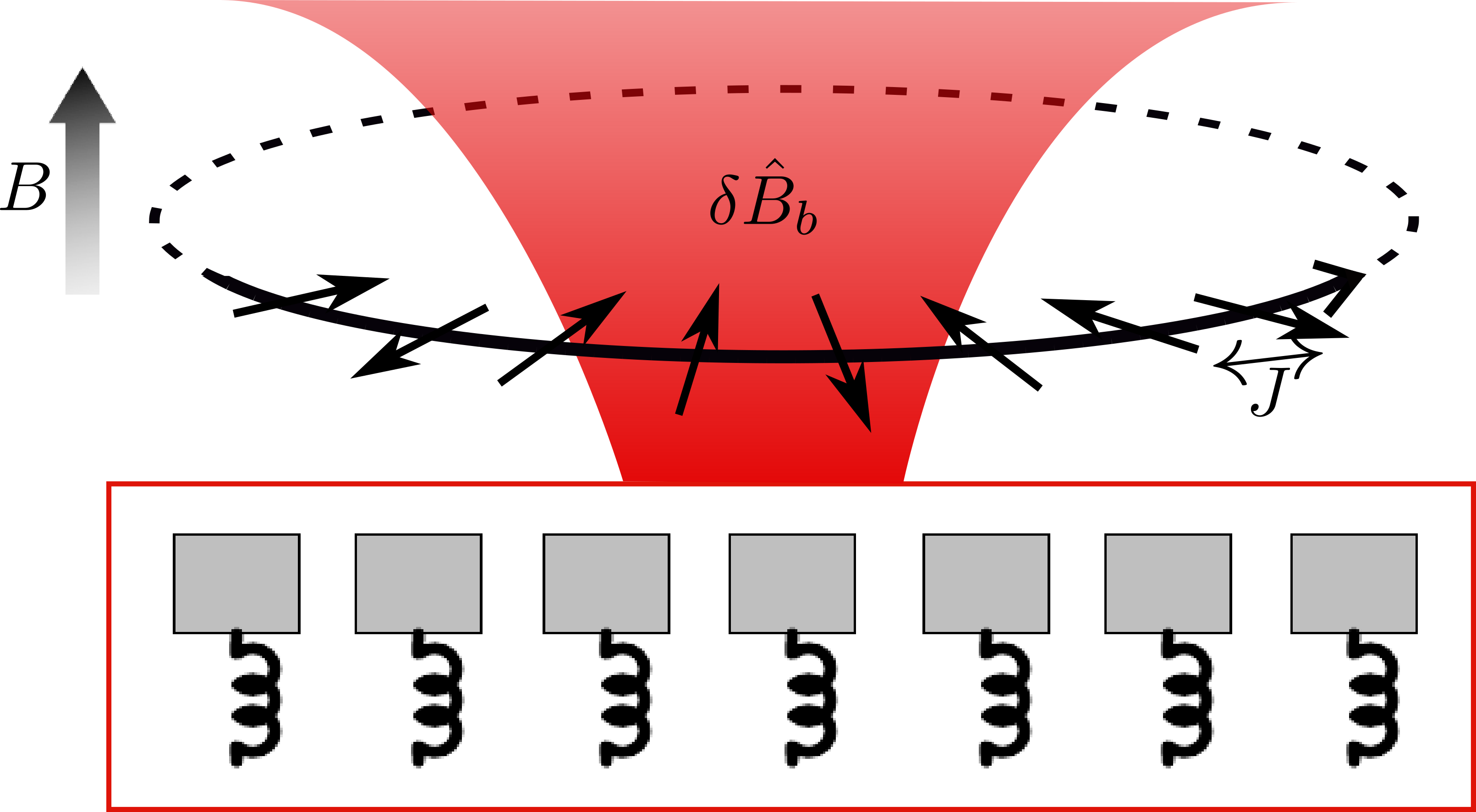}
\caption{A quantum Ising chain forming a ring. 
The interaction among spin, whose strength coupling is $J$, 
is along the $x$ direction, perpendicular the external, transverse magnetic  field $B$. 
The spin are uniformly coupled to a single bath which induces fluctuations in 
the transverse field.}
\label{fig:schema_model}
\end{figure}
%
%
%

Beyond the phase diagram, the study of the relaxation and the decoherence dynamics in the open quantum Ising chain, {\sl with  
a finite number of spins}, represents a relevant theoretical issue. 
This study is important since, generally, the decoherence rate scales 
with the system size and this property might limit the scalability 
in quantum computation.
Moreover, the interplay between dissipation and internal interactions in quantum many-body systems gives  rise to interesting phenomena.
For example, interactions lead to a decoherence dynamics which can be characterized by a (slow) algebraic decay\cite{Cai:2013ea},  
in contrast to a (fast) exponential  decrease.
This  time scaling is a consequences  of the vanishing of the gap in the spectrum of the Liouville superoperator.
%
%
%
%
In another example, non-linear interaction between harmonic oscillators can lead to the formation of maximally entangled states which 
are protected against phase-flip noise  \cite{Coto:2016ck}.

In this perspective, dissipative quantum Ising chains represent a benchtest  
to understand  dissipative many-body systems.

The effects of the dissipation  can depend crucially on the operators coupling to the bath. 
In the works \cite{Werner:2005bf,Werner:2005,Cugliandolo:2005,Sperstad:2010,Hoyos:2012fh}, 
each spin was coupled  to the environment through the same spin  direction of the spin-spin interaction. 
However, the dissipative coupling of the 1D Ising chain with the environment can occur even through other directions.
For instance, the chain can have a dissipative coupling along the direction of the transverse magnetic field.
This form of dissipative interaction was considered to address the dynamical phase 
transition \cite{Mostame:2007ig,Patane:2008fc,Yin:2014gj,Nalbach:2015dk}, 
and  the nonequilibrium state of the chain coupled to two  baths at different temperature \cite{Vogl:2012es}.
Transverse dissipative coupling  could also be an asset for quantum optimization  
since, in this model, quantum diffusion  can provide a mechanism of the speed-up  \cite{Smelyanskiy:2017dw}.

In this work we study the  transverse dissipative coupling for a quantum Ising chain in the limit of  
strong interaction strength among the spins, see Fig.~\ref{fig:schema_model}. 

In section \ref{sec:model} we introduce the model.
In section \ref{sec:strong_coupling} we recall the results for the quantum Ising model in strong coupling regime 
for a chain  {\sl of finite size}.
The many body states are eigenstates of the parity, defined in the direction of the transverse field. 
The ground state subspace is degenerate and it is spanned by the two classical ferromagnetic states, whereas 
the first excited subspace is spanned by pairs of domain walls separating two regions of different magnetization.
Owing to the finite transverse magnetic field, the excited states corresponds to the quantum coherent superposition of such classical domain walls: 
they are delocalized in the chain and have  even or odd parity. 
In section \ref{sec:Lindblad}, assuming the Markovian and weak coupling regime with a single bath, 
we derive the Lindblad equation in terms of the ladder operators associated to the spectrum obtained 
in the strong coupling regime.
The Lindblad equation can be further simplified if the thermal bath has finite but sufficiently low temperature with respect to 
coupling strength $J$ which sets the excitation energy scales.
Then the dynamics can be reduced to the subspace of the ground and first excited states.
Since the dissipative interaction preserves the parity, the ladder operators only connect states with the same parity.
For instance, for vanishing temperature, we obtain that the system can relax only to two ground states of fixed and opposite parity, 
corresponding to the symmetric and antisymmetric superposition of the two classical ferromagnetic states.
In section \ref{sec:discussion} we calculate the relaxation rates between the first excited states and the ground 
states in each parity sector. 
We found that, in the limits here assumed, some excited states have vanishing relaxation rate.
We discuss the properties of such relaxation free subspace by using a mapping   
into a tight binding model.
We summarize our results in the last section \ref{sec:conclusions}.

%
%
%
\section{Model: spin lattice with transverse dissipation}
\label{sec:model}

In general, there are different ways to couple a quantum Ising lattice to the environment.
A class of dissipative models is described by the following Hamiltonian 
\begin{equation}
\label{eq:H_tot}
\hat{H}_{\alpha\beta}  = 
B   \sum_{n=0}^{N-1} \hat{\sigma}_n^z 
 - J \sum_{n=0}^{N-1} \hat{\sigma}_n^\alpha \hat{\sigma}_{n+1}^\alpha 
 +  \delta \hat{B}_b \sum_{n=0}^{N-1} \hat{\sigma}_n^{\beta}
 + \hat{H}_{b} \, ,
\end{equation}
depending on two parameters $\alpha,\beta = x,z$.
A summary of all cases is in given in the Table \ref{table}.
Here,  $\hat{\sigma}^{\alpha}_n$  are the the Pauli matrix for $\alpha,\beta=x,y,z$ for the $n$th spin, 
for instance  $[\hat{\sigma}^x_n , \hat{\sigma}^y_n] = 2i  \hat{\sigma}^z_n$.
The first term in Eq.~(\ref{eq:H_tot}) is the  nearest neighbour  interaction among the $N$ spins along the axis $\alpha$, 
assumed ferromagnetic $(J>0)$ to be definite.
We assume a ring geometry, namely the periodic boundary condition $\hat{\sigma}_{N}= \hat{\sigma}_{0}$. 
The second term is the external transversal magnetic field fixed acting on each spin in $z$ direction.
The third term corresponds to the uniform noise operator $\delta \hat{B}_b$ coupled to 
the total spin component $\hat{S}^{\beta}=\sum_{n} \hat{\sigma}_n^{\beta}$ 
of the spin lattice.
The last term  $\hat{H}_b$ is the bath Hamiltonian whose form is not necessary to specify to derive the Lindblad equation.

For $\alpha=\beta=z$, the interaction among the spin and the operator $\hat{S}^{z}$ 
have the same direction of the transverse field. 
If we assume a Caldeira-Leggett model for the bath, corresponding to an 
ensemble of independent harmonic oscillators, with $\delta\hat{B}_b$ equals to the sum of position operators, 
then the model can be exactly solved and the decoherence  dynamics  is  equivalent to the one 
of the non-interacting case $J=0$ (see appendix \ref{app:exact}).
This model was used as a playground  to illustrate the pure dephasing or decoherence regime for a single 
spin  \cite{Breuer:2012,Duan:1998hw}.

Then the interaction and the coupling to the bath can still have  the same direction for $\alpha=\beta=x$
but they can be both orthogonal to the transverse field.
In this situation of {\sl parallel} dissipation, the phase diagram was explored using  quantum 
Montecarlo simulations  \cite{Werner:2005bf,Werner:2005,Cugliandolo:2005,Sperstad:2010}. 
Since the interaction operator and dissipative coupling operator commute, the result in the phase diagram is 
that the stronger the coupling with the environment is, the larger is the ordered phase region (the ferromagnetic phase), 
namely the critical ratio for $J/B$ for the quantum phase transition decreases with the dissipation.

%
%
%
%
%
\begin{table}[t]
\centering
\begin{tabular}{|p{0.3cm}|p{0.9cm}|p{0.8cm}|p{5.8cm}|}
\hline
$B$		  	& Spin  		 		& Dis.    		&  Model \\
		  	& inter. 	 			& inter.  			&   \\
\hline
$Z$  & $Z-Z$	&\hspace{2mm} $Z$		& exactly solvable (e.g. appendix~\ref{app:exact}) \\
\hline
$Z$ & $X-X$	&\hspace{2mm} $X$ 	&  parallel dissipation  (e.g. \cite{Werner:2005})\\	
\hline
$Z$  & $Z-Z$	&\hspace{2mm} $X$ 	& generalized Dicke  model (e.g. \cite{Tolkunov:2007id}]) \\			
\hline
$Z$ & $X-X$	&\hspace{2mm} $Z$		& transverse dissip. (e.g. \cite{Patane:2008fc} + this work)\\
\hline
\end{tabular}
\caption{Summary of the class of dissipative Ising models described by Eq.~(\ref{eq:H_tot}). 
The transverse magnetic field $B$ is along the $z$ axis.
}
\label{table}
\end{table}
%
%

When the interaction among the spins has the same direction of the transverse field $\alpha=z$ but 
perpedicular to the noise $\beta=x$, the system corresponds to a generalized Dicke model whose phase diagram 
has been analyzed in the framework of the Holstein-Primakoff transformation \cite{Tolkunov:2007id}.

Finally, the case $\alpha=x$ and $\beta=z$ is the quantum Ising model with transverse dissipation 
and it has the explicit form 
\begin{equation}
\label{eq:H_work}
\hat{H}_{xz}  = 
B   \sum_{n=0}^{N-1} \hat{\sigma}_n^z 
 - J \sum_{n=0}^{N-1} \hat{\sigma}_n^{x} \hat{\sigma}_{n+1}^{x} 
 +  \delta \hat{B}_b \sum_{n=0}^{N-1} \hat{\sigma}_n^{z}
 + \hat{H}_{b} \, .
\end{equation}
In this work we analyze the model Eq.~(\ref{eq:H_work}) 
%
which can represent, for instance, a
chain of coupled qubits with equal, individual frequency of $2B$.
Then the interaction with the bath describes a pure dephasing coupling.
%

%
%
%
\section{Strong coupling approximation for the spin interaction}
\label{sec:strong_coupling}

Before to analyze  the effects of the  transverse dissipation, 
we first discuss the quantum Ising model in absence of the coupling with the bath.
We focus on the strong coupling regime $J \gg B$.
This regime can be also realized starting from the situation $J \ll B$ and using a weak driving for the spins 
(see appendix \ref{app:RWA}).

\begin{figure}[b]
\includegraphics[width=\columnwidth]{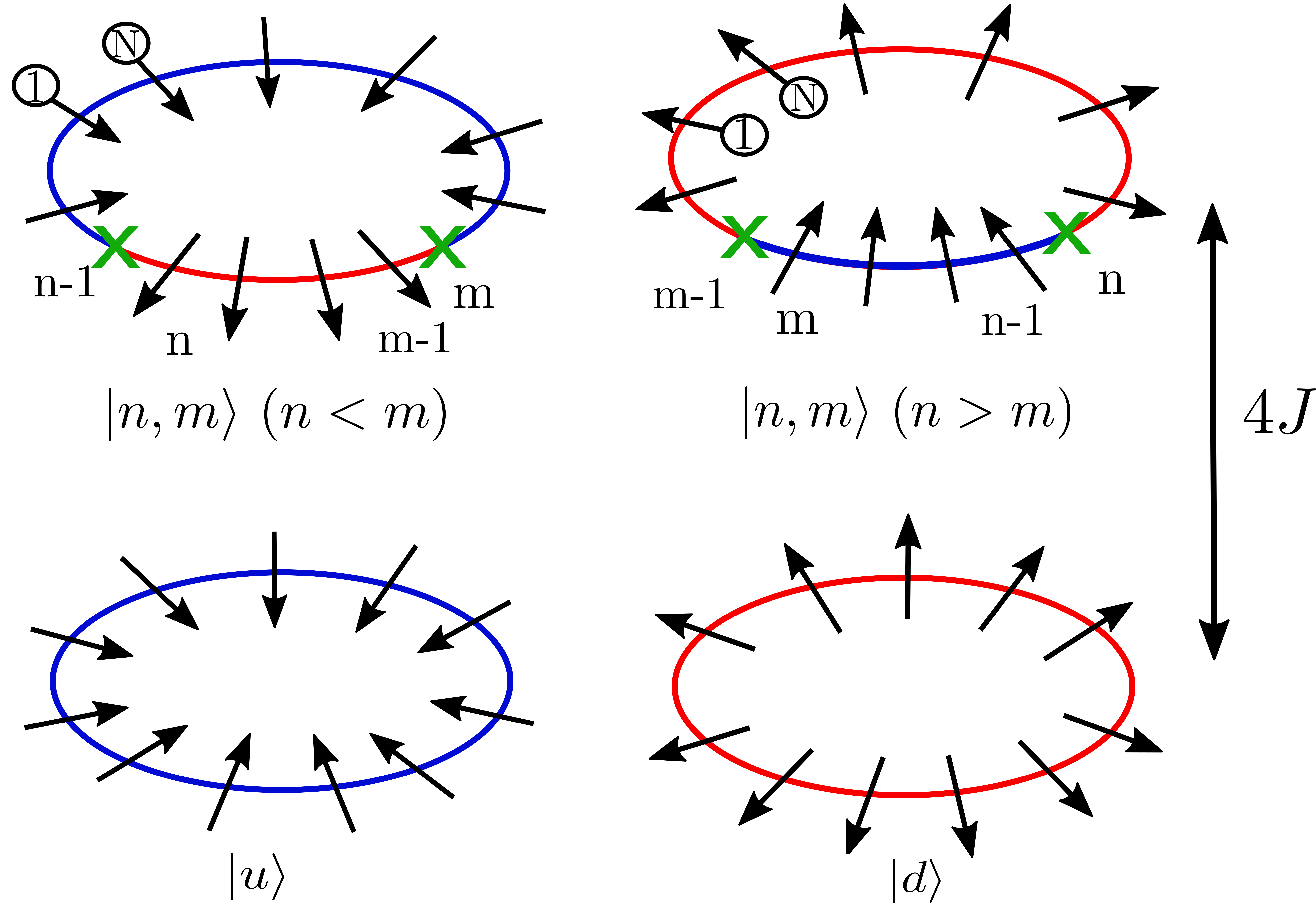}
\caption{
In the limit $B=0$,  the ground subspace of the quantum Ising model is double degenerate and it is spanned   by two degenerate ferromagnetic 
states  $\ket{u}$  and  $\ket{d}$.
The first excited subspace is formed by  single pairs or domain walls (kinks) at energy $\Delta E=4 J$.
The index $n$ and $m$ are the position of the domain walls, $n$ for the domain wall from an up-spin area 
(blue) to a down-spin area (red) and $m$ for the reversed case.
}
\label{Fig.spinrings}
\end{figure}

Starting from the limit  $B=0$, the spectrum of the transverse Ising model assumes a simple form.
The ground state subspace is double degenerate and it is spanned by the two ferromagnetic 
states {\sl along the $x$ direction}, Fig.~\ref{Fig.spinrings}.
The excited subspaces are formed by states  corresponding to domain walls separating  regions 
of parallel spins with different direction.
For instance, the first excited subspace contains only one pair of domain walls (see Fig.~\ref{Fig.spinrings}), 
it has energy $4J$ respect to the ground state with degeneracy $N(N-1)$. 
%
%
In the first excited subspace, we denote the state of two pairs as $  \ket{n,m}  $ in which the first index refers to the  position of the 
domain wall between $n$ and $n+1$  where the $x$ spin component  changes from up to down, 
whereas the second index refers to the position domain wall  
between $m$ and $m+1$, where the $x$ spin component  changes from down to top, 
see Fig.~\ref{Fig.spinrings}.
Notice that two states of inverted indeces are different $  \ket{n,m}  \neq  \ket{m,n} $.

%
%
%
%
%
%
%
%

\begin{figure}[t]
\includegraphics[width=\columnwidth]{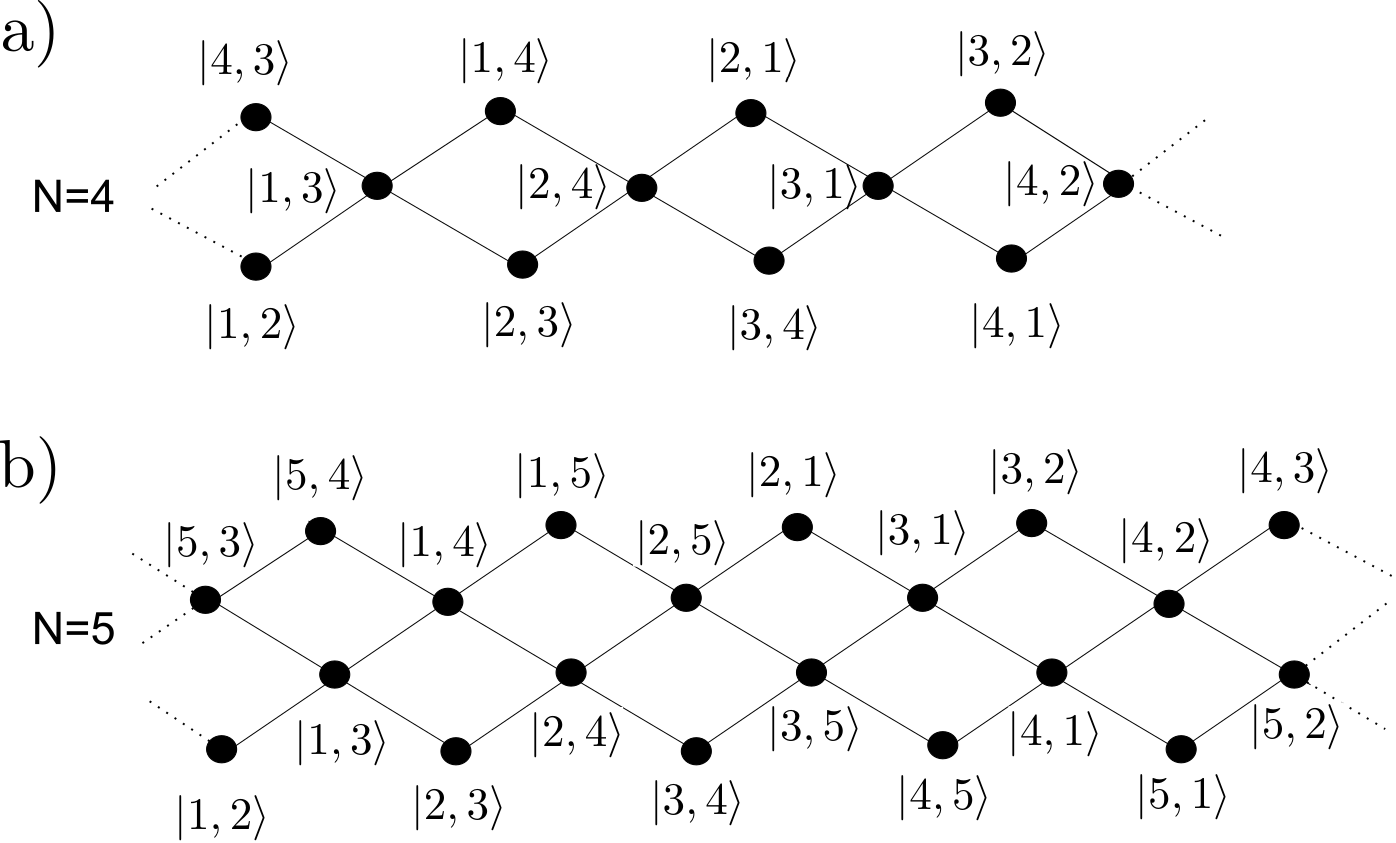}
\caption{
The states $\ket{n,m}$ of the first excited subspace in the limit of $B=0$
are represented on an effective lattice for  the case (a) $N=4$  and (b) $N=5$.
Restoring finite $B\ll J$, the perturbation operator in this subspace  $\hat{S}^{z}_1$  acts as hopping operator 
connecting these states according the effective lattice.
States on the borders  corresponds to states with kinks separation of one.
Dotted lines represent periodic boundary conditions.
}
\label{Fig.lattice}
\end{figure}

Restoring the finite value of $B$, in the limit  $B \ll J$, 
one applies the perturbation theory for the degenerate case by diagonalizing 
the states within each excited subspace with respect to the perturbation operator $\hat{S}^z$.
This removes the degeneracy within the excited subspaces.
If the interaction with the bath (dissipation) is not too strong (see next discussion), 
we can consider only the first excited subspace composed only by one pair of domain walls, i.e. 
the states $\ket{n,m}$.
Setting the projection operator $\hat{P}_1$
on the first excited subspace, we can write  $\hat{S}^z_{1}= \hat{P}_1\left( \sum_n \hat{\sigma}^z_n \right) \hat{P}_1$ 
and the eigenstates of this operator in the first excited subspace are simply given by 
$\hat{S}^z_{1} \ket{\Psi} = \varepsilon \ket{\Psi}$, which can be expanded in the basis of  $ \ket{n,m} $ 
\begin{equation}
\label{eq:Psi}
\ket{\Psi}= \!\!\! \sum_{n,m=1}^N  f_{(n,m)}   
\ket{n,m}  =  \!  \sum_{n=1}^N\sum_{m=n+1}^{N+n-1} f_{(n,m)}  \ket{n,m}, 
\end{equation}
with the condition $f_{(n=m)}=0$.
In the second equality (\ref{eq:Psi}) we introduced the periodic notation $\ket{n,m} = \ket{n+N,m} =\ket{n,m+N}$.
In the first excited subspace, the transverse magnetic field coupling acts similar to the hopping operator of the tight binding model 
\begin{equation}
\label{eq:Szp_1}
\hat{S}^z_{1}\ket{n,m}   =  \sum_{s=-1,+1} \ket{n+s,m}  +   \sum_{s=-1,+1} \ket{n,m+s}  \, ,
\end{equation}
for $|n-m|>1$ whereas for $|n-m|=1$ or  $|n-m|=N-1$ we have 
\begin{align}
\hat{S}^z_{p}   \ket{n,n+1} &=    \ket{n,n+2}  + \ket{n-1,n+1}  \, , \label{eq:Szp_2} \\
\hat{S}^z_{p}   \ket{n,n-1} &=    \ket{n,n-2}  + \ket{n+1,n-1}  \, . \label{eq:Szp_3}
\end{align}
The tight binding Eqs.~(\ref{eq:Szp_1}),(\ref{eq:Szp_2}) and (\ref{eq:Szp_3}) 
are associated to an effective lattice for the states $ \ket{n,m} $ whose examples 
are reported in Fig.~(\ref{Fig.lattice}) for $N=4$ and $N=5$.
This effective lattice is a closed ribbon with periodic boundary condition only in one direction,  as indicated with the dotted lines, 
and open boundaries in the second direction. 
In general, the ribbon has the (periodic) perimeter $2N$ and finite size width  $N-1$.
States with kinks distance of one site, e.g. $\ket{n,n\pm1}$, are on the 
borders and they have only two connections as given by Eqs.(\ref{eq:Szp_2},\ref{eq:Szp_3}).
The corresponding Schr{\"o}dinger equation for the wavefunction $f_{n,m}$ can be solved 
and  the spectrum is described by two quantum numbers $k$ and $q$ 
\begin{equation}
\varepsilon(k,q) = 4 B \cos\left(\pi k/N \right )    \cos\left( \pi q / N \right )  \, ,
\end{equation}
with the eigenstates  
\begin{equation}
\label{eq:ket_kq}
\ket{k,q} \!\! = \!\! 
\frac{\sqrt{2}}{N} \!\!\!
\sum_{n,m=1}^N 
\!\! 
\sin\left[ \frac{\pi}{N}k(m-n) \right] e^{i\frac{\pi}{N}q(m+n)}  \xi_{(n,m)}^{(k,q)}  
\ket{n,m}  
\end{equation}
with $ \xi_{(n,m)}^{(k,q)}  = 1$ for $n<m$ and $ \xi_{(n,m)}^{(k,q)}  = e^{i\pi(k+q)}$ for $n>m$ 
(see appendix \ref{sec:eigenstates} for details).

%
%
%
%
%
%
%
%

\section{Derivation of the Lindblad equation in the strong coupling}
\label{sec:Lindblad}

Generally, when the spin lattice interacts with the bath via the total spin component $\hat{S}^z$, an exact solution is not possible.
Here we analyze the problem within the Lindblad equation. 

Under the assumption of a weak dissipative interaction between the system with the bath, in the Markovian regime 
(the Born-Markov approximation) and using a rotating wave approximation, 
the equation of the reduced density matrix  $\hat{\rho}$
of a quantum, open systems, in the interacting picture in respect to the bath, reads \cite{Breuer:2012} 
\begin{equation}
\frac{d \hat{\rho}(t)}{dt}=  - \sum_{\omega_i}  \,  \gamma(\omega_i) \,\,  \hat{\mathcal{L}}_{\omega_i} [\hat{\rho}(t)] \, , 
\end{equation}
where we omitted the Lamb-shift renormalization of the spectrum. 
The function $\gamma(\omega)$ is related to the symmetrized noise of the bath operator $\delta\hat{B}$, namely 
$\gamma (\omega) =  \int_0^\infty \!\! dt  e^{i \omega t}  {\langle  \{ \delta\hat{B}_b(t) , \delta\hat{B}_b(0) \}  \rangle}_{b} $,
where the $\delta\hat{B}_b(t)$ refers to the free evolution and ${\langle \dots \rangle}_b$ the trace on the thermal state of the bath
(see also appendix \ref{app:exact} for the notation).
This function describes the processes of emission or absorption of the energy of the bath and it can be written as  
\begin{equation}
\gamma (\omega) =
K(\omega) \left[  \theta(\omega)  \left( 1+n_B (\omega) \right)+  \theta(-\omega) n_B (-\omega) \right]  \, .
\end{equation} 
where $K(\omega)$ (defined as even function) is the spectral environmental  interaction function \cite{Breuer:2012}. 
The superoperator $\hat{\mathcal{L}}_{\omega_i} $ acts as 
\begin{equation}
\hat{\mathcal{L}}_{\omega_i} [\hat{X}]
= 
\frac{1}{2} 
\left\{    
\hat{A}^\dagger(\omega_i)  
\hat{A}(\omega_i) ;  \hat{X}  
\right\}  
-
\hat{A}(\omega_i)  \hat{X}   \hat{A}^\dagger(\omega_i) 
\end{equation}
with  $\hat{A}(\omega_i)$ corresponding to the ladder operators  
\begin{equation}
\label{eq:ladders}
\hat{A}(\omega_i)
=  
\sum_{E_{\alpha},E_{\beta}} \, \delta_{\omega_i, E_{\alpha}- E_{\beta}} \, 
\hat{\Pi}(E_{\beta})  \hat{S}^z   \hat{\Pi}(E_{\alpha}) \, .
\end{equation}
Here ${E_{\alpha}}$ is the energy level spectrum of the system $\hat{H}_s$ and the operator $\hat{\Pi}(E_{\alpha})$ 
is the projector on the  (degenerate) subspace of energy $E_{\alpha}$.
Compared to the case of the individual coherent systems \cite{Breuer:2012}, here the spectrum corresponds the 
the many-body states of the quantum Ising lattice. 
We remark that our approach is different from other theoretical Lindblad models  
in which the ladder operators are expressed in term of single spin operators.
%
%

The ladder operators connect the eigenstates of the spectrum formed by the ground states 
and the excited subspaces. 
In the limit of small transverse magnetic field, the energy spacing is given by the coupling strength $J$ 
(see Fig.~\ref{Fig.spinrings}). 
To further proceed, we assume that the symmetrized noise decreases with the energy spacing, such as  
$\gamma(4J(n+1))  \ll \gamma (4Jn) $, and we can restrict  the Lindblad equation  to 
the two ground states and the first excited subspace whose spectrum reads $E_1(k,q) = E_{GS} + 4 J + \varepsilon(k,q)$.

We first discuss the action of the ladder operators within each energy subspace.
The projectors on the ground state subspace, with $\omega_i=0$, has zero matrix element in  Eq.~(\ref{eq:ladders}) 
since $\hat{S}^z$ flips always one spin of the lattice.
The projectors on the first excited subspace in  Eq.~(\ref{eq:ladders})  which has energy spacing $|\omega_i|=\varepsilon(k,q)-\varepsilon(k',q')$ a priori, 
give a finite contribution only for $\omega_i=0$ ($\varepsilon(k,q)=\varepsilon(k',q')$) since the first excited subspace is formed 
by the eigenstates of the operator $\hat{S}^z_1$ with $\hat{S}^z_1 \equiv \hat{S}^z$  and we obtain 
\begin{equation}
\label{eq:A_0}
\hat{A}(\omega_i=0) =  \hat{A}_0 =\hat{A}^{\dagger}_0 = \sum_{k,q} \frac{\varepsilon(k,q)}{B}    \ket{k,q} \bra{k,q}  
\end{equation}
Second, we have to discuss the ladder operators Eq.~(\ref{eq:ladders}) connecting the ground states with the first excited subspace 
which are characterized a priori by  energy spacing $|\omega_i | = 4J + \varepsilon(k,q) \simeq 4J$.
However, since $\hat{S}^z$ can flip at most one spin, one can create only domain walls of distance one 
by applying $\hat{S}^z$ to one of the two ferromagnetic ground state states (Fig.~\ref{Fig.spinrings}), i.e. 
$\hat{S}^z$ has non zero matrix element between the ground states and  $\ket{n,m}$ with $|n-m|=1$ or $|n-m|=N$. 
In the effective tight-binding lattice representation (Fig.~\ref{Fig.lattice}), these states are located on the borders.
This condition sets the following matrix element rule: only the states $\ket{k,q}$ with quantum number $q=N$ have 
non-vanishing matrix element of the operator  $\hat{S}^z$ with the ground state.
Setting
\begin{equation}
E_k = 4J + \varepsilon(k,N)  \, ,
\end{equation}
 we denote the ladder operators  
\begin{equation}
\hat{A}(\omega_i = E_k) \equiv \hat{A}_{k,s} \,\,\,\, , 
\hat{A}(\omega_i = -E_k) = \hat{A}^{\dagger}_{k,s} \, .
\end{equation}
Notice that a global sign of the operator $\hat{A}_{k,s}$ is irrelevant. 
The explicit form of the operator $\hat{A}_{k,s}$ is given by  
\begin{equation}
\label{eq:A_k}
\hat{A}_{k,s} = 2 \sin  \left(  \frac{\pi k}{N}   \right)  \,\, \delta_{s s_k}  \,\, s_{k} \ket{s}  \bra{k,N}  \, , 
\end{equation}
where $s=\pm$ and $s_{k} = \exp[i \pi (k+N+1)]$.
The state $\ket{s=+}$ and $\ket{s=-}$  are the two ground states corresponding to the 
symmetric and anti-symmetric linear combination  of the two ferromagnetic ground states $\ket{u}$ and $\ket{d}$
\begin{equation}
\label{eq:g_k}
\ket{\pm} =  \frac{1}{ \sqrt{2} } \left( \ket{u}    \pm  \ket{d} \right)  \, .
\end{equation}
The results Eq.~(\ref{eq:A_k})  and Eq.~(\ref{eq:g_k}) have a simple physical explanation.
The states $\ket{\pm} $  correspond  to the two ground states which are also eigenstates of the parity operator  defined as 
\begin{equation}
\mathcal{P} = \prod_n \hat{\sigma}_n^z \, .
\end{equation}
The states $\ket{k,N}$ are also eigenstates of the parity operator 
and the Lindblad operator Eq.~(\ref{eq:A_k}) connects states with equal parity.
For instance, for $N$ odd, the state $\ket{+}=\ket{u}    +   \ket{d}$ is connected 
only with the states of  $k$ even whereas the state $\ket{+}=\ket{u}   -    \ket{d}$
is connected  with the state of  $k$ odd, namely the phase factor $ e^{i\pi(k+N+1)}= \pm 1$, 
The situation is inverted for a lattice of even length $N$.

Finally, collecting the previous results Eq.(\ref{eq:A_0}) and  Eq.(\ref{eq:A_k}),
we obtain the expression for the Lindblad equation for the quantum Ising lattice in the strong coupling regime and 
with transverse dissipation in the limit in which higher energy subspaces are neglected 
\begin{align}
\frac{d\hat{\rho}}{dt} 
&\simeq -\gamma(0) \left[  \frac{1}{2} \left( \hat{A}^2_0 \hat{\rho} +\hat{\rho} \hat{A}^2_0  \right) - \hat{A}_0\hat{\rho}\hat{A}_0   \right]  \nonumber \\
- &  \gamma\left(4J \right) \sum_k \left[ \frac{1}{2} \left( \hat{A}_k  \hat{A}_k^{\dagger} \hat{\rho} +\hat{\rho} \hat{A}_k  \hat{A}_k^{\dagger}  \right) 
-  \hat{A}_k^{\dagger} \ \hat{\rho} \hat{A}_k  \right] \nonumber \\
- &  \gamma\left(-4J\right)  \sum_k \left[ \frac{1}{2}  \left( \hat{A}_k^{\dagger}  \hat{A}_k \hat{\rho} +\hat{\rho} \hat{A}_k^{\dagger}  \hat{A}_k \right) 
-  \hat{A}_k  \ \hat{\rho} \hat{A}_k^{\dagger}  \right]  \label{eq:Lindblad} \, ,
\end{align}
in which we have omitted the index $s$ referring to each parity subspace.

%
%
%
%
%
\begin{figure}[t]
\includegraphics[width=\columnwidth]{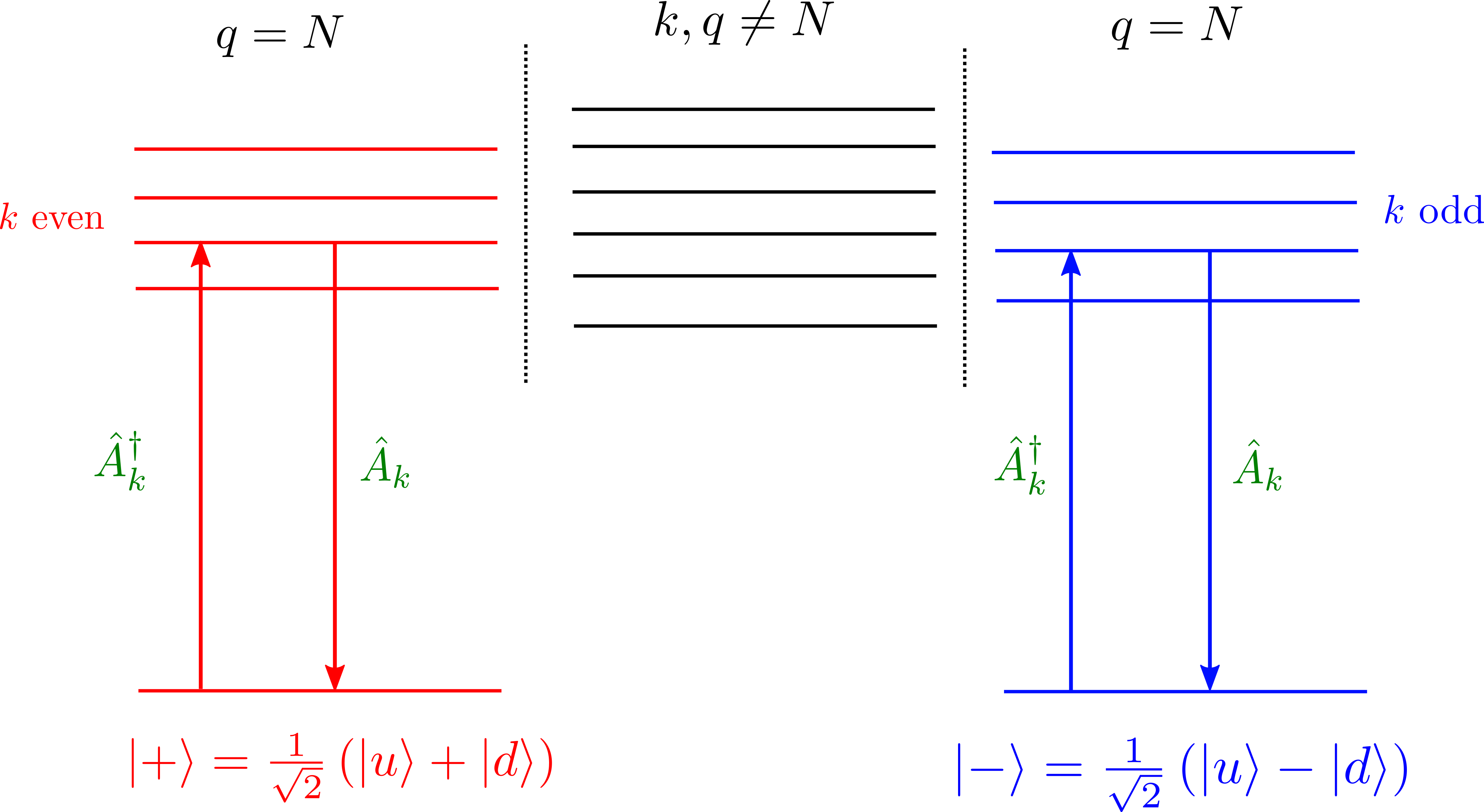}
\caption{Schematic picture for the ground states and first excited subspace whose eigenstates are described by the quantum numbers $k$ and $q$.
The parity is preserved under the effect of the dissipative transverse interaction with the bath.
The ladders operators appearing in the Lindblad equation are shown by arrows and they connected the ground states with defined parity 
to the excited states of equal parity and quantum number $q=N$.}
\label{Fig.ladders}
\end{figure}
%
%
%

\section{Discussion and results of the Lindblad equation}
\label{sec:discussion}

The Lindlblad Eq.~(\ref{eq:Lindblad})  governs the quantum dissipative dynamics of the system.
For the rest, it is worthy to distinguish between the three different sets: (i) $\mathcal{E}_g$ the two ground states  
($\ket{\pm}$) (ii) $\mathcal{E}_{N}$ the states in the first excited subspace with quantum number $q=N$ (with $k$ even or odd), 
and (iii) $\mathcal{E}_{\tilde{q}}$ the states in the first excited subspace with quantum number $q \neq N$ 
(see Fig.~\ref{Fig.ladders}).
Using these sets of states, it is possible to show that the Lindlblad Eq.~(\ref{eq:Lindblad})   takes a simple form.
First, the off-diagonal element of the density matrix are completely decoupled from the diagonal elements and 
they satisfy a close equation, i.e. $\dot{\rho}_{\lambda,\lambda'} = - (1/\tau_{\lambda,\lambda'}) \rho_{\lambda,\lambda'}$
for two arbitrary different states $\lambda \neq \lambda'$.
Second, the diagonal elements of the density matrix, or populations, of the ground state space $\mathcal{E}_g$
and the excited subspace $\mathcal{E}_{N}$  are the only ones which are coupled whereas the populations 
of the subspace $\mathcal{E}_{\tilde{q}}$ are decoupled from them (see Fig.~\ref{Fig.ladders}).

\subsection{Populations for the $\mathcal{E}_g$  and $\mathcal{E}_{N}$  states}
More precisely, we have a set of equations  in each parity sector.
Hereafter, to be define, we set $N$ odd.
Setting $P_{+} = \rho_{++}$, and $P_{k}= \rho_{kk}$ for the populations of the states $\mathcal{E}_{k}$ with $k=k_e$ even, 
we have 
\begin{equation}
\label{eq:Ppdot}
\frac{d P_+}{dt}
= \sum_{k_e} 
- 4\sin^2 \left(\frac{\pi}{N}k_e \right) \left[  \gamma(4J) \, P_{+}  \, -  \, \gamma(-4J) \, P_{k_e}  \right] \\
\end{equation}
\begin{equation}
\label{eq:Pkdot}
\frac{d P_{k_e} }{dt}
=
- 4 \sin^2 \left( \frac{\pi}{N}k_e \right) \left[  \gamma(4J)  \, P_{k_e}  -  \gamma(-4J)  P_{+}  \right] 
\end{equation}
and similar equations for the population $P_{-} = \rho_{--}$, and $P_{k}= \rho_{kk}$  
for the populations of the states $\mathcal{E}_{k}$ with $k=k_o$ odd.

As mentioned before the conserved quantity is the parity, 
thus the sums of the populations at time $t=0$ in the even 
sector $P_e =P_{+} + \sum_{k_e} P_{k_e}$
and in the odd sector  $P_o =P_{-} + \sum_{k_o} P_{k_o}$ are invariant during the decay.
%
%
 By using the  detailed balance relation $\gamma(-E)/\gamma(E) = e^{-\beta E}$, 
 the steady state solutions of the coupled equations Eq.~(\ref{eq:Ppdot}) and Eq.~(\ref{eq:Pkdot}) read 
\begin{align}
P_{+}      & = \, P_{e}  \,  \frac{1}{ 1+ \mathcal{N}_{e}   e^{ -4\beta J} }				\\
P_{k_{e}} & = \, P_{e}  \, \frac{  e^{ -4\beta J} \ }{ 1+ \mathcal{N}_{e}  e^{ -4\beta J}   }
\end{align}
with $\mathcal{N}_{e}$ the number of the even $k$ states in  $\mathcal{E}_{k}$. 
Similar expressions hold for the odd sector, $P_{-}$ and $P_{k_{o}}$ with the constant $P_o$ and number  $\mathcal{N}_{o}$.

The exponential relaxation  associated to the coupled equations Eq.~(\ref{eq:Ppdot}) and Eq.~(\ref{eq:Pkdot})  
is determined by the characteristic relaxation time.
As example, we discuss the low temperature limit and set $k_B T \ll 4J$ such that $\gamma(-4J) \simeq 0$ exponentially.
For the even sector we have the following exponential decays 
\begin{align}
P_{+}(t)     &= \, \sum_{k_e} \, P_{k_e}(0)  \,  \left( 1 - e^{- t /\tau_{k_e} } \right) \\
P_{k_e}(t)	&=	\, P_{k_e}(0) \, e^{- t /\tau(k_e) }
\end{align}
and similar expressions for the odd sector, for $P_{-}(t)$ and $P_{k_o}(t)$. 
The relaxation times are given by 
\begin{equation}
\frac{1}{\tau(k)} = 4K(4J) \sin^2\left(\frac{\pi}{N}k\right)	\, ,
\end{equation}
and the states with $k\approx N/2$ are more rapidly decaying as the states with $k$ close to $N-1$ or  to $1$.
In particular, the lowest relaxation rates scales as $\tau_k^{-1} \sim {(\pi k / N)}^2$ and 
it is strongly reduced in the limit  $N \gg 1$ and $k \ll N$.

\subsection{Relaxation-free subspace $q \neq N$}

As mentioned before, the diagonal elements of the density matrix for the states $\ket{k,q}$ with $q \neq N$ are 
decoupled from the population for the ground states and the states with $q=N$.
This implies that, within the limit here discussed, these states do not relax to the ground state.
Notice that decoherence among these states is still possible and eventually, the density matrix will converge to a statistical mixture 
of these state $\ket{k,q\neq N}$.

%
%
%
%
%
\begin{figure}[t]
\includegraphics[width=\columnwidth]{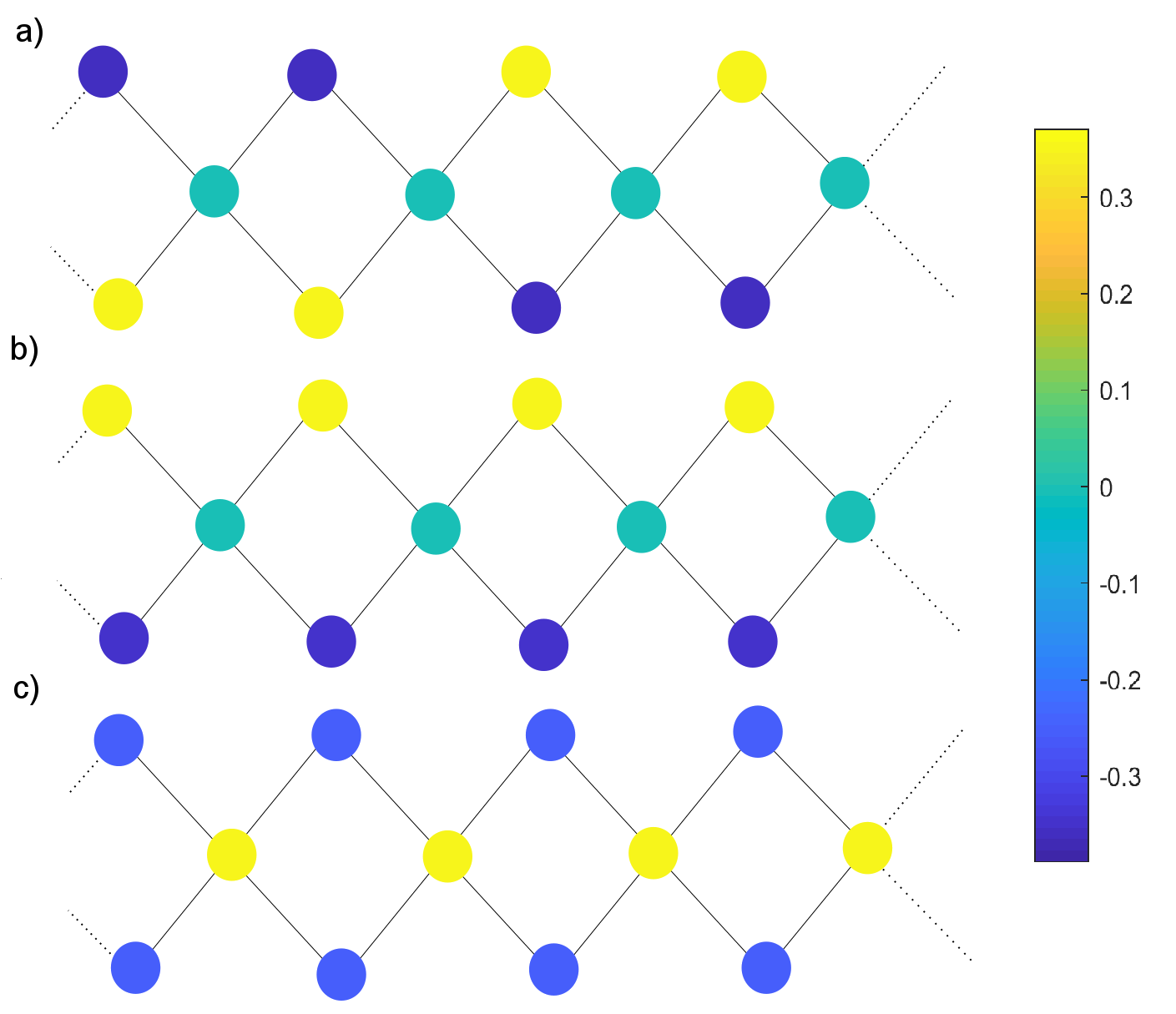}
\caption{Example of the wavefunctions of the eigenstates 
$ \ket{k,q} = \sum_{n,m} f_{n,m}(k,q) \ket{n,m}$
of the first excited subspace in the limit $B \ll J$, defined on the effective lattice, shown in Fig.\ref{Fig.lattice}, 
for the kink states $\ket{n,m}$.  
For $N=4$, we plot $\mbox{Re}[f_{n,m}(k,q)]$ .
(a) is an example of the states uncoupled to the ground state subspace, $\ket{k=2,q=3}$.
(b) and (c) are examples of the states coupled to the ground state subspace, with $\ket{k=2,q=N}$ and  $\ket{k=1,q=N}$, 
respectively.
}
\label{Fig.modes}
\end{figure}
%
%
%

Considering the Eq.~(\ref{eq:ket_kq}), one can observe that the index $q$ plays the role of the wavevector associated to the 
center of mass of the kink pair $\sim n+m$ whereas $k$ plays the role of the wavevector related to the internal distance $\sim n-m$ 
(see appendix \ref{sec:eigenstates} for details).
In this way we can interpret the states $q=N$ as coherent superposition of different bound states of kinks of different distance 
and with vanishing total momentum.
More precisely the condition for decoupling from the relaxation with the ground states is given by
\begin{equation}
\label{eq:condition}
\langle \left. k,q \right| \hat{S}^z \ket{u} = \langle\left. k,q \right| \hat{S}^z \ket{d}  = 0  \, .
\end{equation}
The condition Eq.~(\ref{eq:condition}) is equivalent to the condition  that the sum of the amplitudes corresponding to the states of 
distance one as $\ket{n,n+1} $  (the states on one boundary in the tight binding lattice, see Fig.~\ref{Fig.lattice})  
and $\ket{n,n-1}$  (the states on the opposite boundary in the tight binding lattice, see Fig.~\ref{Fig.lattice})  
has to vanish. 
Illustrating examples  are reported in Fig.~\ref{Fig.modes}.
For instance, the state $\ket{k=2,q=3}$ is not coupled to the ground state subspace 
and its wavefunction has an antisymmetric behaviour on the boundaries of the lattice. 
Another example is the state $\ket{k=2,q=N}$ which is  coupled to the ground state $\ket{-}$ 
since its coefficients on the two boundaries  have different sign, whereas 
the state $\ket{k=1,q=N}$ (b) is coupled to the ground  state $\ket{+}$ 
since  its coefficients on the two boundaries  have the same sign.

\section{Summary}
\label{sec:conclusions}
To summarize, we studied a quantum Ising lattice dissipatively coupled to a single bath which 
describes the quantum fluctuations of the transverse magnetic field. 
In the limit of strong coupling among the spins, we discussed the excitations of the first excited subspace 
corresponding to coherent quantum superposition of the classical states formed by pairs of domain walls  
that divides  regions of different ferromagnetic ordering.
Using Born-Markov approximation, we derived an effective Lindblad equation with the ladder operators 
associated to the transition between the many-body states of the lattice, i.e. the ground states and the excitations 
of quantum kinks. 
The parity symmetry of the quantum Ising model, defined with respect to the axis of the transverse field, 
is still conserved in presence of transverse dissipation.
This implies that, in the Lindblad equation, the two sectors of different parity are decoupled and 
the excited states can relax only to the ground state of equal parity.
We also found that, in the limits here assumed, different states excitations can not relax to the ground state 
and they form a relaxation-free subspace.
These states are characterized by a quantum number $q \neq N$ and they can be interpreted 
as the states {\sl with vanishing total amplitude on the borders} of an effective tight binding lattice 
formed by the single pair kink  at different places $\ket{n,m}$. 
A such result motivates future extensions of our analysis to decorated lattices 
with more complex interaction beyond the Ising model.

%
%
%
%
%
\acknowledgments
This work was supported by the German Excellence Initiative through the Zukunftskolleg, 
the Deutsche Forschung Gemeinschaft  (DFG) through the SFB 767, 
and by the MWK-RiSC program. 
%
%
%
%
%

%
%
%
\appendix

%
%
%
%
\section{Exactly solvable case}
\label{app:exact}

The model Hamiltonian Eq.~(\ref{eq:H_tot}) for a dissipative spin chain with  $\alpha=\beta=z$ reads 
\begin{equation}
\hat{H}_{zz}  =  \hat{H}_s + \hat{H}_{I}   + \hat{H}_{b}  \, ,
\end{equation}
where we set
\begin{align}
\hat{H}_s  &=\sum_{n} \left(  B - J \hat{\sigma}_{n+1}^{z} \right) \hat{\sigma}_n^z \, , \\
\hat{H}_{I}   &=  \delta \hat{B}_b  \, \hat{S}^z   \, ,
\end{align}
with $\hat{S}^z  = \sum_{n} \hat{\sigma}_n^{z}$, 
can be solved exactly for arbitrary boundary conditions, namely a ring or an open chain. 
Following the Caldeira-Leggett approach, we assume the bath as a sum of independent harmonic oscillator
$\hat{H}_{b}  = \sum_{\lambda} \hbar \omega_{\lambda} \hat{b}^{\dagger}_{\lambda} \hat{b}^{\phantom{g}}_{\lambda}$, 
$\delta\hat{B}_b =  \sum_{\lambda} \alpha_{\lambda} (  \hat{b}^{\dagger}_{\lambda} +  \hat{b}^{\phantom{g}}_{\lambda} )$ 
and the bosonic operator (creation and annihilation) $ \hat{b}^{\phantom{g}}_{\lambda}$ and $\hat{b}^{\dagger}_{\lambda} $.
Assuming the free evolution of this operator $\hat{b}_{\lambda}(t)= e^{i \hat{H}_b t} \hat{b}_{\lambda} e^{-i \hat{H}_b t}$, 
the commutator at different times for $\delta\hat{B}_s$  (the response function) simplifies to 
\begin{equation}
\left[ \delta \hat{B}_b(t_1) , \delta \hat{B}_b (t_2) \right] 
=  -2i
\int_0^\infty \!\!\!\!\!\!\! d\omega  K(\omega)  \sin\left[ \omega (t_1-t_2)\right] \, ,
\end{equation}
where we introduced the environmental spectral density $K(\omega) = \sum_{\lambda} \alpha^2_{\lambda} \delta(\omega-\omega_{\lambda})$
In similar way, the average at thermal equilibrium of the  anticommutator at different times for $\delta\hat{B}_s$  
(the symmetrized noise) reads
\begin{equation}
{\left<  \left\{  \delta \hat{B}_b(t_1) , \delta \hat{B}_b (t_2) \right\}  \right>}_{b} 
\!\!\!\!
=  
\!\!
\int_0^\infty \!\!\!\!\!\!\! d\omega K(\omega)\! \coth\left(\!\frac{\beta \omega}{2}\!\right) \!\! \cos\left[ \omega (t_1\!-\!t_2)\right]
\end{equation}
with average over the thermal state of the bath ${\left<   \dots  \right>}_{b} = \mbox{Tr}_b [e^{- \beta \hat{H}_b} \dots]  /\mbox{Tr}_b[e^{-\beta \hat{H}_b}]$, 
and $\beta=\hbar/(k_B T) $.
For the following calculations, it is useful to introduce the two functions
\begin{align}
\phi (t) &=  -i  \int_0^t \!\!\!\! dt_1  \int_0^{t_1} \!\!\!\!  dt_2 \,   \left[  \delta \hat{B}_b(t_1) , \delta \hat{B}_b (t_2)  \right] \, , \\
\zeta(t) &= \int_0^t \!\!\!\! dt_1  \int_0^{t_1} \!\!\!\!  dt_2  {\left<   \left\{ \delta \hat{B}_b(t_1) , \delta \hat{B}_b (t_2)  \right\}  \right>}_{b}  \, .
\end{align}
In the interaction picture $\hat{H}_I(t) = e^{i (\hat{H}_s + \hat{H}_b ) t}  \hat{H}_I  e^{-i (\hat{H}_s + \hat{H}_b ) t}$,
the unitary time evolution operator $\hat{U}_I(t) = \hat{T} e^{-i \int_0^t  dt'  \hat{H}_I(t)} $ ($\hat{T}$ time ordering operator)  
can be calculated exactly
\begin{equation}
\hat{U}_I(t) = 
e^{ - \frac{i}{2} \hat{S}^{z}  \hat{S}^{z}   \phi(t) } \,\, 
e^{-i  \hat{S}^{z}  \int_0^t  dt'  \delta\hat{B}_n(t) }
\end{equation}
In this way, we can calculate the reduced density matrix of the system $H_S$
\begin{equation}
\rho_s(t)= \mbox{Tr}_b \left(    \hat{U}_I(t)  \rho_S(0) \rho_b   U_I^\dagger(t)  \right)  
\end{equation}
in which  $\rho_s(0) $ the initial state of the system and $\rho_b$ the thermal state of the bath.
For the eigenstates of the spin lattice we use the notation  $\ket{\left\{s\right\}}=\ket{s_1,\dots,s_n,\dots}$ 
with $s_n = \pm 1$ for the spin component up or down in the $z$ direction, and the energies 
$E_s = \sum_n ( B - J s_{n+1} ) s_n$ and total $z$ componet $\hat{S}^z \ket{\left\{s\right\}} = ( \sum_{n} s_n)  \ket{\left\{s\right\}}$. 
After some algebra, similar to the case of non-interacting spins \cite{Breuer:2012}, 
the arbitrary matrix element of the reduced density matrix  $ \rho_{ss'}(t)= \bra{  \left\{s\right\} }   \rho_S(t)   \ket{ \left\{s\right\}'  } $ 
can be written as 
\begin{equation}
\left| \rho_{ss'}(t)  \right| =  \left| \rho_{ss'}(0)    \right| 
e^{- \left( \sum_{n} s_n - \sum_{n} s^{'}_n \right)^2   \zeta(t)} \, .
\end{equation}
The last result is exactly the same for the case $J=0$ \cite{Breuer:2012}.

%
%
%
%
\section{Driven spin lattice in Rotating Wave Approximation}
\label{app:RWA}
A weak external driving of the spins can be a possible way to reach the strong coupling regime discussed in the main text 
\begin{align}
\hat{H}_s(t) & = \sum_{n} \left(  B \hat{\sigma}_n^z - J \hat{\sigma}_{n+1}^{x} \hat{\sigma}_n^x \right)   
+
\sum_{n}F_{n}\left(\hat{\sigma}_{n}^{+} e^{i\omega_{d}t} + h.c. \right) \, .
\end{align}
Transforming into the rotating frame with $\hat{U}_{d} =e^{-i\frac{1}{2} \omega_{d} t \hat{S}^z }$ and using the 
Rotating Wave Approximation (RWA), we obtain 
\begin{align}
\hat{U}_{d}^{\phantom{g}}  \hat{\sigma}_{n}^{x} \hat{\sigma}_{n+1}^{x} \hat{U}_{d}^{\dagger} 
& \simeq \left(\hat{\sigma}_{n}^{x}\hat{\sigma}_{n+1}^{x}+\hat{\sigma}_{n}^{y}\hat{\sigma}_{n+1}^{y}\right) / 2
\end{align}
in which we neglect fast oscillating terms.
The transformed Hamiltonian in the rotating frame $H^{'} = 
\hat{U}_{d}^{\phantom{g}}  \hat{H}_s   \hat{U}_{d}^{\dagger}  -  i  \hbar \hat{U}_{d}^{\phantom{g}} 
\partial \hat{U}_{d}^{\dagger} / \partial t$ in the RWA reads 
\begin{equation}
\hat{H}_{R} 
\simeq 
(B-\hbar\omega_{d})\sum_{n}\sigma_{n}^{z}
-  J \sum_{n}\left(\hat{\sigma}_{n}^{x}\hat{\sigma}_{n+1}^{x}+\hat{\sigma}_{n}^{y}\hat{\sigma}_{n+1}^{y}\right)  / 2
\end{equation}
For weak driving $|F_{n}| \ll |B- \hbar \omega_{d}|$ the term $\sum_{n}F_{n} \hat{\sigma}_n^{x}$ can be neglected.
By a rotation of angle $\pi/4$ in the xy-plane, one recovers
\begin{equation}
\label{eq:H_RWA}
J\sum_{n}\left(\hat{\sigma}_{n}^{x}\hat{\sigma}_{n+1}^{x}+\hat{\sigma}_{n}^{y}\hat{\sigma}_{n+1}^{y}\right) / 2 =J
\sum_{n}\left(\hat{\sigma}_{n}^{x'}\hat{\sigma}_{n+1}^{x'}\right) /\sqrt{2} \, .
\end{equation}
The Eq.~(\ref{eq:H_RWA}) describes a quantum Ising Hamiltonian with the effective magnetic field  
$B^{'}=B-\hbar\omega_{d}$.
By varying $\omega_D$, one can reach  the strong coupling regime $|B-\hbar \omega_{d}| \ll J$.

\section{Solution for the eigenstates of the first excited subspace  - quantum delocalized kinks}
\label{sec:eigenstates}
From the tight binding Eqs.(\ref{eq:Szp_1},\ref{eq:Szp_2},\ref{eq:Szp_3}) 
we obtain to the following set of equations:
\begin{equation}
B \left[ f_{(n+1,m)}  + f_{(n-1,m)} + f_{(n,m+1)} + f_{(n,m-1)} \right] = 
\varepsilon f_{(n,m)} \label{eq_fnm} \, , 
\end{equation}
for $|n-m|>1$ whereas we have 
\begin{align}
B \left[ f_{(n+1,m)}  + f_{(n,m-1)} \right] = \varepsilon f_{(n,m)}   \,\,  &\mbox{for} \,  n=m+1 \, , \label{eq_fnm_I}  \\
B \left[ f_{(n-1,m)}  + f_{(n,m+1)} \right] = \varepsilon f_{(n,m)}  \,\, &\mbox{for} \,\, n=m-1 \, , \label{eq_fnm_II}  \, .
\end{align}
%
%
%
%
Notice that the Eqs.(\ref{eq_fnm_I}),(\ref{eq_fnm_II}) are automatically included in Eq.(\ref{eq_fnm}) within the condition 
$f_{(n=m)}=0$.
We represent this set of equations in term of a tight-binding model in which 
the lattice sites are associated to the eigenstates $\ket{n,m}$ for a pair of kinks 
and the lines between two sites represent the connection (hopping) between two states due to the operator $\hat{S}^z$
(see examples in Fig.\ref{Fig.lattice}). 
It is convenient to define the directions
\begin{align}
x&=
\begin{cases}
n+m& $for $n<m\\
n+m\pm N& $for $ n>m
\end{cases}
\label{eq:x_direction}
\\
y&=
\begin{cases}
m-n& $for $n<m\\
m-n+N& $for $ n>m \, , \label{eq:y_direction}
\end{cases}
\end{align}
such that  
the tight binding equation remains invariant  
\begin{equation}
B \left(f_{x+1,y+1}+f_{x+1,y-1}+f_{x-1,y-1}+f_{x-1,y+1}\right) = \epsilon f_{x,y} \, ,
\end{equation}
with $f_{x=y}=0$. 
Them the boundary conditions in the $x$ direction are periodic whereas we have open boundary conditions in the $y$ direction  
\begin{align}
f_{x+2N,y}=f_{x,y}& \, , \quad \quad f_{x,N}=f_{x,0}=0 \, .
\end{align} 
The latter equations suggest  the ansatz
\begin{align}
f_{x,y}^{(k,q)}& \sim  \sin\left(  \frac{\pi}{N} k y  \right) \, e^{i \frac{\pi}{N} q x}  \, , 
\end{align}
for $k=1,...,N-1$ and $q=1,..,N$, which yields the spectrum
\begin{align}
\frac{ \varepsilon(k,q) } {2 B} &=  \cos\left[ \frac{\pi}{N} \left( k - q \right) \right] +  \cos\left[ \frac{\pi}{N} \left( k + q \right)  \right]\, .
\end{align}
However the two coordinates $x$ and $y$ can be written simply as the centre of mass and the relative distance of the two kinks function 
($x \sim n+m$ and $y \sim n-m$) only for $n<m$ and one needs to consider the extra factor $N$ for $n>m$ as explained in 
Eq.~(\ref{eq:x_direction}) and  Eq.~(\ref{eq:y_direction}).
Taking into account this difference, the exact form for the eigenstates reads 
\begin{align}
f_{x,y}^{(k,q)}& \sim   \sin\left(  \frac{\pi}{N} k [m-n]  \right) \, e^{i \frac{\pi}{N} q (n+m)}  \, \mbox{for} \,\, n<m \\
f_{x,y}^{(k,q)}& \sim   \sin\left(  \frac{\pi}{N} k [m-n + N]  \right) \, e^{i \frac{\pi}{N} q (n+m\pm N)} \nonumber \\
&
\sim \sin\left(  \frac{\pi}{N} k [m-n]  \right) \, e^{i \frac{\pi}{N} q (n+m)} e^{i\pi(k+q)}
 \, \mbox{for} \,\, n>m \, ,
\end{align}
that corresponds to the (unnormalized) result reported in the main text, Eq.~(\ref{eq:ket_kq}).

\bibliography{bibliography}

\end{document}